\documentclass[conference]{IEEEtran}
\IEEEoverridecommandlockouts
\usepackage{cite}
\usepackage{amsmath,amssymb,amsfonts}
\usepackage{algorithmic}
\usepackage{graphicx}
\usepackage{textcomp}
\usepackage{xcolor}
\usepackage{hyperref}
\def\BibTeX{{\rm B\kern-.05em{\sc i\kern-.025em b}\kern-.08em
    T\kern-.1667em\lower.7ex\hbox{E}\kern-.125emX}}
\begin{document}

\title{A sinusoidal signal reconstruction method for the inversion of the mel-spectrogram}

\author{\IEEEauthorblockN{Anastasia Natsiou}
\IEEEauthorblockA{\textit{Technological University of Dublin} \\
Dublin, Ireland \\
anastasia.natsiou@tudublin.ie}
\and
\IEEEauthorblockN{Se\'{a}n O'Leary}
\IEEEauthorblockA{\textit{Technological University of Dublin} \\
Dublin, Ireland \\
sean.oleary@tudublin.ie}
}

\maketitle

\begin{abstract}
The synthesis of sound via deep learning methods has recently received much attention. Some problems for deep learning approaches to sound synthesis relate to the amount of data needed to specify an audio signal and the necessity of preserving both the long and short time coherence of the synthesised signal. Visual time-frequency representations such as the log-mel-spectrogram have gained in popularity. The log-mel-spectrogram is a perceptually informed representation of audio that greatly compresses the amount of information required for the description of the sound. However, because of this compression, this representation is not directly invertible. Both signal processing and machine learning techniques have previously been applied to the inversion of the log-mel-spectrogram but they both caused audible distortions in the synthesised sounds due to issues of temporal and spectral coherence. In this paper, we outline the application of a sinusoidal model to the ‘inversion’ of the log-mel-spectrogram for pitched musical instrument sounds outperforming state-of-the-art deep learning methods. The approach could be later used as a general decoding step from spectral to time intervals in neural applications.

\end{abstract}

\begin{IEEEkeywords}
Sound reconstruction, mel-spectrogram, sinusoidal model, machine learning.
\end{IEEEkeywords}

\section{Introduction}
Digitised audio signals can be represented in different forms depending on the application and the available resources. \textit{Raw audio} is a representation of the time domain acoustic signal specified by the sampling rate and bit depth. This form is typically the end goal of sound synthesis algorithms. However, as we perceive sound in terms of time evolving spectral content, time-frequency representations are often appropriate. Such representations include the \textit{spectrogram} based on \textit{short time Fourier transform} (STFT), and the \textit{log-mel-spectrogram}, which is a compressed form of spectrogram modelled after the frequency resolution of the human ear.

During the last years, many different approaches have been applied to sound synthesis. Basic signal processing methods, e.g. \textit{concatenative synthesis} \cite{hunt_unit_1996} and \textit{statistical parametric synthesis} \cite{zen_statistical_2009} were very influential in the field, especially in the area of speech synthesis.
More recently, the rise of computational power and deep learning architectures have led to the development of high quality results for sound synthesis. WaveNet \cite{oord_wavenet_2016} is a state-of-the-art vocoder\footnote{Vocoder comes from the term `voice coder' although it is also used for more general analysis/synthesis models e.g. `phase vocoder'. Here we are using it in the latter more general sense.} based on dilated convolutional neural networks. WaveNet's high quality depends on synthesis directly in the time domain; every new sample is based on previous ones. Therefore, the system is computationally expensive and it needs a large amount of data in order to be trained. 

Many deep learning algorithms have employed spectrogram like representations \cite{prenger_waveglow_2018}\cite{engel_gansynth_2019}\cite{kumar_melgan_2019}\cite{kong_diffwave_2021}\cite{vasquez_melnet_2019}. A major issue with this representation is that it is not directly invertible as it only describes the general spectral distribution over time. In this paper we use the log-mel-spectrogram to generate parameters for a signal processing synthesis model. We implemented a modification of the classic sinusoidal model generating amplitudes that approximate the spectral distribution of the log-mel-spectrogram; the phase calculation takes advantage of the continuous nature of quasi-sinusoidal oscillations over time. Thus, our system overcomes some of the issues of deep learning based inversion, such as that used in Tacotron 2 \cite{shen_natural_2018}, being a less computationally expensive decoding step and shows the potential of producing higher quality results for the class of sounds under consideration. 

The paper is organized as follows. In Section 2 we review previous works on sound reconstruction methods, and in Section 3 we describe the sinusoidal model and our modification for the inversion of the log-mel-spectrogram. Results are presented in Section 4, and conclusions in Section 5.

\section{Related Work}
In many visual representations of sound, the phase part of a time-frequency representation is discarded. This is generally appropriate for sustained sounds as the ear is not sensitive to absolute phase. To reconstruct a pitched waveform from its given "phaseless" spectrogram, many signal processing-based and machine learning-based methods have been proposed.

Signal processing methods for sound generation focus directly on attempting to reconstruct the phase from the spectrogram. A much used solution for phase estimation is the Griffin-Lim algorithm \cite{griffin_signal_1984} \cite{griffin_new_1985}. This method manages to estimate the parameters by minimizing a least square error criterion between the generated and the original spectrum. The algorithm is able to attain impressive results and it is used by many researchers even today \cite{wang_tacotron_2017}. Another significant vocoder for generating sound from a time-frequency spectrogram is STRAIGHT \cite{kawahara_restructuring_1999}. The key objective of this algorithm is to eliminate the periodicity of a speech signal and then use a sampling operation to generate the parameters. WORLD \cite{morise_world_2016} is designed to reconstruct the sound by its fundamental frequency, the spectral envelope and an aperiodic parameter. Due to its quick processing and its low computational cost, WORLD is able to generate sound with adequate quality in real time. 

Most of the models lack of sound naturalness, producing a "robotic" effect. Consequently, machine learning has been adopted widely. WaveNet \cite{oord_wavenet_2016} is an autoregressive and probabilistic neural network-based architecture applied on raw audio. Although WaveNet was initially proposed as a stand-alone application, in Tacotron 2 \cite{shen_natural_2018}, it was conditioned on log-mel-spectrogram and used as a vocoder. Another implementation of deep learning for the inversion of the log-mel-spectrogram is WaveGlow \cite{prenger_waveglow_2018}. This method models multiple invertible transformations to estimate and sample from the distribution of the training data. The generated sound achieved demonstrates the same quality as WaveNet but with less computational cost.

To take advantage of the strengths of both methodologies, some contemporary attempts include hybrid models. LPCNet \cite{valin_lpcnet_2019} was proposed as an improvement to the WaveRNN \cite{kalchbrenner_efficient_2018}. A Linear Prediction Coefficient analysis on Bark-scale cepstral coefficients was applied to generate the training data of a neural network. This way, they managed to model the response of the vocal tract and improve the synthesised sound.

Although our approach clearly falls into the signal processing category, our short term plan includes the investment to a more hybrid method, using the method as a general decoding step of a neural architecture; an approach conceptually similar to the recent DDSP \cite{engel_ddsp_2020}. In DDSP, a differentiable autoencoder has been applied for the parameter estimation in a supervised or unsupervised manner. The model manages to synthesise monophonic sounds using the fundamental frequency, the loudness and a latent representation which forms the timbre. The computed parameters are then fed to a harmonic oscillator and a noise filtering component.

\section{Sinusoidal Model}
Although deep learning models demonstrated some results, their demand for a large amount of data as well as their high computational cost lead researchers to prior signal processing or hybrid methods. The sinusoidal model \cite{mcaulay_speech_1986} constitutes a signal processing method that provides a parametric approximation of a signal. In the following section, after introducing the mel-scale, we will explain the original sinusoidal model, and then we will apply the effective inversion of the log-mel-spectrogram.

\subsection{The mel-spectrogram}
In 1937, a group of psychologists conducted a psychoacoustical experiment concerning the perceived distance in pitch between sounds with varying frequency \cite{volkmann_scale_nodate}. The outcome was that the human ear has greater resolution at lower frequencies. This observation led to the development of the \textit{mel-scale} (mel from 'melody'). The conversion of the frequency in hertz (f) to the mel-scale is illustrated in Eq.\ref{eq_mel}.

\begin{equation}
    mel = 2595log_{10}(1+\frac{f}{700})
    \label{eq_mel}
\end{equation}

The mel-spectrogram demonstrates a compressed form of sound in the time-frequency domain. This nonlinear transformation constitutes the outcome of the Short Time Fourier Transform (STFT) after the application of mel-filters (a bank of bandpass filters with bandwidths modelled after the mel-scale).

\subsection{Sinusoidal Model from Linear Spectrogram}

The sinusoidal model \cite{mcaulay_speech_1986} models signals as a linear combination of sinusoids, each having independent time varying amplitude and phase trajectories. The modeled signal is given by
\begin{equation}
\label{eqn_1}
y\left(t\right) = \sum\limits_{l=1}^{L} A_{s}\left(t\right)\cos \left(\theta_{s}\left(t\right)\right)
\end{equation}
where $\theta_{s}(t)  = \int_0^t 2\pi f_l (\tau)\,d\tau  + \phi_l\ $is the time evolving phase of the $s_{th}$ sinusoid \footnote{Equation \ref{eqn_1} is in continuous time, what follows will be in reference to frames of the STFT}.

The basic steps of analysis include a \textit{peak detection} of the spectral shape in each frame of the STFT, a \textit{parameter estimation} using parabolic interpolation, and a \textit{peak matching} across frames of the SFTF. The synthesis is then the addition of the sinusoids modelled from the time evolving trajectories.

\subsection{Sinusoidal Model from Mel-Spectrogram}

The following method uses the log-mel-spectrogram and an estimate of the fundamental frequency to synthesise the target sound. The log-mel-spectrogram constitutes a non-invertible representation since many frequencies can co-exist in the same mel-band. Furthermore, the mel-spectrogram encodes only the energy distribution over time-frequency of the given sound, therefore, it represents a phaseless representation. In this paper, we propose a sinusoidal model to model a harmonic source, using the log-mel-spectrogram to inform the spectral distribution. While not attempting to estimate the absolute phase of the original signal, this will preserve continuity in the oscillations of the signal. The approach outlined here can be seen as a source/filter approach - shaping a spectrally flat source with a time variant filter.  

\subsubsection{Frequency Estimation}
For monophonic harmonic target signals, the sinusoid partials are described as integer multiples of the fundamental frequency \cite{beauchamp_analysis_2007}. 
Therefore, the harmonic frequencies can be calculated according to the Eq. \ref{freqs} where $f_{n}^{1}$ is the fundamental frequency of the $n$\textit{th} frame, i=[$1, floor\left[\frac{f_{s}/2}{f_{n}^{1}}\right] $].
 
\begin{equation}
    f_{m}^{i}=if_{m}^{1}
    \label{freqs}
\end{equation}
where $f_{m}^{i}$ is the frequency of the $i$\textit{th} sinusoid in the $m$\textit{th} frame and $f_{s}$ is the sampling rate.

In addition, in order to guarantee no distortions of the synthesised signal based on fluctuations of the fundamental frequency, a 6\% tolerance is adopted between adjacent frames.

\subsubsection{Amplitude Computation}
To estimate the amplitude of each sinusoid, the energy of every frame of the log-mel-spectrogram needs to be preserved. Conversely, having the mel-spectrogram and the information of the mel-filter banks, the amplitude of each sinusoid can be approximated using a source filter model. As this project demonstrates an initial attempt of this idea, a filter equivalent to the rectangular one is applied. Finally, because a single frequency can be present in more than one filter, the amplitude computed is the average of those estimated in overlapping filters.

\subsubsection{Phase Reconstruction}
The phase of each estimated frequency peak is calculated by the phase of the previous frame and the frequencies on both the previous and the current frame interpolated by the hope size as it is expressed in Eq. \ref{phase}. 

\begin{equation}
    \theta_{n}^{i}=\theta_{n-1}^{i}+2\pi T\frac{f_{n}^{i}+f_{n}^{i-1}}{2}
    \label{phase}
\end{equation}
where $\theta_{n}^{i}$ is the phase of the $i$\textit{th} sinusoid in the $n$\textit{th} frame, $T$ is the hop size multiplied by $f_{s}$.

\section{Experiments}
To evaluate the performance of the modified sinusoidal model, we examined the inversion of the mel-spectrogram for independent pitches of a variety of instruments. The results are compared to a pre-trained WaveNet conditioned on the log-mel-spectrogram and to a pre-trained WaveNet autoencoder \cite{engel_neural_2017}. The sounds used in this experiment can be found here: \url{https://anastasianat57.github.io/sinusoidalreconstruction/}.

\subsection{Dataset}
For the conducted experiments, we used a subsample of the NSynth dataset \cite{noauthor_nsynth_nodate}, trying to be as diverse as possible. For each instrument, three different music notes were captured; a low, a medium and a high pitch, in a range of 80Hz to 2100Hz. The sounds could belong in different categories as per their velocity or acoustic quality.

\subsection{Description of the experiments}
The sinusoidal model approach as well as WaveNet are utilised in this paper in order to reconstruct the original music notes from their the log-mel-spectrogram. As we explained before, the models are based on two completely different technologies, therefore the comparison occurs between signal processing and deep learning approaches independently for the conversion of a frequency representation to time domain. The log-mel-spectrogram was created after the raw audio of the original sound with 80 bins, originating from an STFT of 1024 window length, 256 hop size and 1024 FFT size. More details about the parameters of both models are given in the following sections.

\subsubsection{WaveNet}
The sinusoidal model is compared with an already pre-trained WaveNet on speech \cite{noauthor_open_nodate}. These results can be considered as baseline because both the models are conditioned on the same log-mel-spectrogram.

\subsubsection{WaveNet Autoencoder}
The main comparison can be considered between our approach and a WaveNet autoencoder pre-trained on the NSynth dataset \cite{noauthor_magentamagenta_nodate}. Although the network is not conditioned on the log-mel-spectrogram, the original tests \cite{engel_neural_2017} demonstrated that the representation generated from the autoencoder can outperform a baseline using spectrograms. However, we did not consider their baseline as an interesting comparison to our method.

\subsubsection{Sinusoidal Model}
For the sinusoidal model, a normalized blackman window is used to address the expectations of the amplitude computation and the yin algorithm searches for fundamental frequencies between 80Hz and 3000Hz. 

\subsection{Evaluation}

After reconstructing the sample from its original representation, the models are evaluated using spectral convergence. For this purpose, the synthesised sounds are aligned with the original music note and the error is calculated by the Eq. \ref{mse}, where $S$ represents the power spectrogram of the original, $\widehat{S}$ the power spectrogram of the synthesised and $n$ the FFT size of the power spectrogram.

\begin{equation}
    SC = \sqrt{\frac{\sum_{i=1}^{m}\sum_{j=1}^{n}|S(i,j)-\widehat{S}(i,j)|}{n+m}}
    \label{mse}
\end{equation}

Finally, the two aligned sounds are normalized by a total energy calculated using the energy of the original and the energy of the generated sound as it is demonstrated in the Eq. \ref{energy}.

\begin{equation}
    energy=\sqrt{\frac{\sum_{i=1}^{m}S(i,j)}{\sum_{i=1}^{m}\widehat{S}(i,j)}}
    \label{energy}
\end{equation}

\subsection{Results}

\begin{figure}\centering
\includegraphics[width=1.0\linewidth]{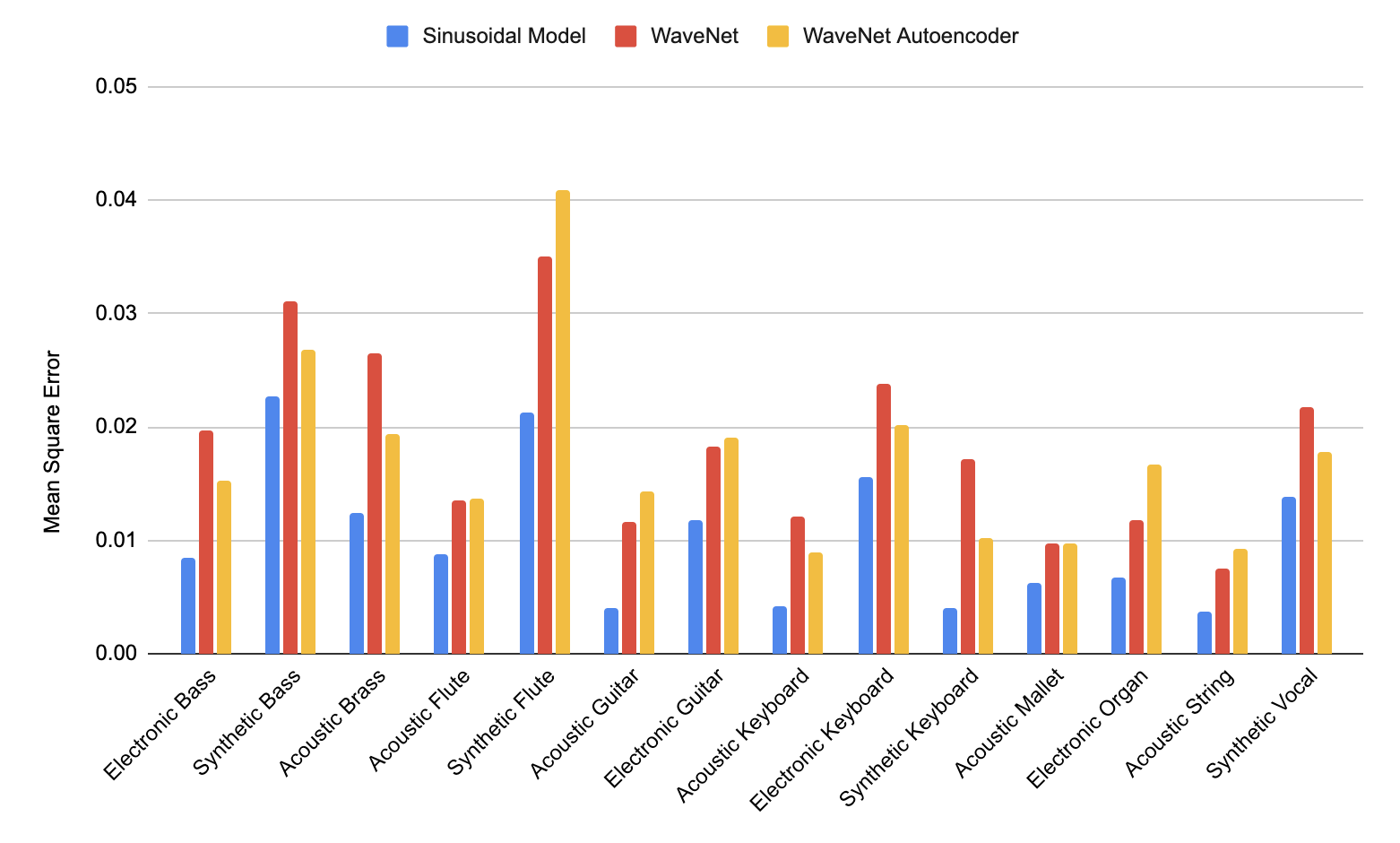}
\caption{Mean Square error of the power spectrogram between the original and the generated sound for both sinusoidal model and WaveNet}
\label{results}
\end{figure}

The results, aggregating them by instrument, are illustrated in Fig. \ref{results}. Using the spectral convergence, one can identify that the modified sinusoidal model is more precise than the WaveNet autoencoder in all the cases. We encourage the reader to listen to the samples provided. The musical note synthesised by the sinusoidal model sounds more natural, presenting phase coherence.

Another advantage of the modified sinusoidal model over deep learning approaches is the low computational cost. The synthesis of a four second musical note using a pre-trained WaveNet model needs significantly more time than the synthesis of the same sound using the sinusoidal model on an average laptop. In cases where the model needs to be trained again, the memory needed and the computational cost exceed the capabilities of an average laptop.

\section{Conclusion}
Deep learning architectures have been proven very effective in sound synthesis applications. WaveNet is a state-of-the-art signal reconstruction model based on dilated convolutional neural networks and it has been used for the inversion of the log-mel-spectrogram, a non-invertible audio representation. While deep learning models do show promise, they can be difficult to interpret and are computationally inefficient at the final synthesis step. Comparing it with a signal processing method, WaveNet autoencoder presents lack of continuity and phase coherence while the requirements for computational power are severe. In the current paper, we proposed a signal processing method based on a modified sinusoidal model for the inversion of the log-mel-spectrogram. The results seem promising, achieving more natural synthesised musical notes than WaveNet with a significantly lower computational cost. Although the outcome of the sinusoidal model illustrated high quality, the current version of the model is not appropriate for the generation of non harmonic signals. With the suitable extraction of the parameters, this model could be used for any sound. This research work illustrates an early investigation of marrying signal processing with deep learning. An extension of the presented model could be used as the last step of a deep learning architecture for decoding spectrogram-based representations for a more general class of single source sounds.

\section*{Acknowledgment}

This work was funded by Science Foundation Ireland through the SFI Centre for Research Training in Machine Learning (18/CRT/6183)

\bibliographystyle{ieeetr}
\bibliography{ref}
\end{document}